\documentclass[conference]{IEEEtran}
\IEEEoverridecommandlockouts
\usepackage{algorithmic}
\usepackage{amsmath,amssymb,amsfonts}
\usepackage{cite}
\usepackage{graphicx}
\usepackage{multirow}
\usepackage{textcomp}
\usepackage{xcolor}

\def\BibTeX{{\rm B\kern-.05em{\sc i\kern-.025em b}\kern-.08em
    T\kern-.1667em\lower.7ex\hbox{E}\kern-.125emX}}

\begin{document}

\makeatletter
\newcommand{\linebreakand}{
	\end{@IEEEauthorhalign}
	\hfill\mbox{}\par
	\mbox{}\hfill\begin{@IEEEauthorhalign}
}
\makeatother

\title{Algorithmic support of a personal virtual assistant for automating the processing of client requests}

\author{
	\IEEEauthorblockN{Konstantin Dobratulin}
	\IEEEauthorblockA{
		\textit{College of Information Technologies} \\
		\textit{and Computer Sciences} \\
		\textit{NUST MISIS} \\
		Moscow, Russia \\
		dobratulin@yahoo.com
	}
	\and
	\IEEEauthorblockN{Marina Nezhurina}
	\IEEEauthorblockA{
		\textit{College of Information Technologies} \\
		\textit{and Computer Sciences} \\
		\textit{NUST MISIS} \\
		Moscow, Russia \\
		min@misis.ru
	}
}

\maketitle

\begin{abstract}
	This article describes creating algorithmic support for the functioning of a
	personal virtual assistant, which allows automating the processing of customer
	requests. The study aims to reduce errors and processing time for a client request
	in business systems -- text chats or voice channels using a text transcription
	system. The results of the development of algorithmic support and an assessment of
	the quality of work on synthetic data presented.
\end{abstract}


\section{Introduction} \label{Introduction}

Personal virtual assistants have gone through a development path of several decades --
from text recognition and transcription systems to multifunctional software agents using
technological advances in artificial intelligence \cite{c01}. In 2022, personal virtual
assistants perform tasks for the end user and are used in business services, such as:
virtual operators, automatic response systems, smart devices, voice assistants,
interactive voice menu \cite{c02}. Personal virtual assistants focused on the end user
are being introduced into software systems of business systems. Using personal virtual
assistants in systems for automated processing of client requests allows businesses to
build new user interaction scenarios, improve business metrics based on resource
optimization and influence on technological, temporal, economic and behavioral factors
\cite{c03}. Such possibilities of use served as a rationale for the need for algorithmic
support for a personal virtual assistant.

Thus, the technological connection of personal devices, multimedia devices and virtual
assistants has opened great opportunities for business: there are integrations with
music and video services, Internet sites, mobile applications. Personal virtual
assistants gained an image, character, and they tried to bring interaction with them
closer to interaction with a person.

The development of personal virtual assistants has made it possible to expand the range
of possibilities for using assistants in solving business problems, to ensure
communication between business and the end user as if communication were built between
two people. The performance of not only routine daily tasks, but also complex business
logic, assigned to personal virtual assistants, has created a demand for application in
all areas of business.

\section{Virtual assistant in request processing systems} \label{Virtual assistant in request processing systems}

Interaction with a virtual assistant can be carried out using various communication
channels, such as sending text requests, voice requests, downloading images or a video
stream \cite{c04}. However, speaking about the processing of requests in an automated
request processing system, we abstract from the request, delegating the transformation
of information coming from the end user into a textual and indicative form. For what
tasks can a virtual assistant be used? The answer to this question appears based on the
needs of the business and the vision of the product in which the virtual assistant will
be used. An increase in business metrics creates criteria that a personal virtual
assistant must meet, what functionality it must have, and what tasks it must provide. Of
course, the first requirement for a modern personal virtual assistant is effective
interaction with the end user, receiving and processing requests. Such basic
functionality can already be adapted to solve typical tasks, and the system itself is
scaled to match the loads in terms of the number of requests in an automated system.
Thus, a list of basic application possibilities appears:

\begin{enumerate}
	\item Reception of user requests in the communication channel of the business area.
	\item Processing and transformation of a user request into a standardized machine-understandable form.
	\item Passing the converted request to the associated software system.
	\item Sending a response to the end user in the business area communication channel.
\end{enumerate}

Such basic capabilities of a personal virtual assistant can already be applied to
communication with the end user, transferring all interaction into the communication
channel required by the business.

Thus, the concept of using a personal virtual assistant in a system for automated
processing of client requests implies the interaction of the end user with a certain
software agent that performs the functionality of the system’s business logic to
effectively perform business tasks in a specific area of activity, while the
functionality of a personal virtual assistant is limited only by the vision of the
product.

\section{Approaches to client data classification for algorithmic support } \label{Approaches to client data classification for algorithmic support }

Linguistic features are an important criterion for classifying customer data. Since the
specifics of interaction with a personal virtual assistant implies that the user sends a
request in a text or voice channel, it becomes possible at the request processing stage
to extract features from the textual representation of the user’s request. Such signs
can be lexical, textual, and lexico-grammatical signs. According to such criteria, it
becomes possible to classify user data, for example, by assigning a certain query
complexity rating based on linguistic features, to analyze the user’s mood based on the
words or phrases used in the query's context, to remove repetitions or insignificant
words by textual features for statistical analysis. When developing a personal virtual
assistant, one should carefully consider the potential of linguistic feature analysis,
as this may require the use of such approaches as those that have become popular today,
approaches from the field of artificial intelligence \cite{c05}.



\section{Modeling the process of obtaining and analyzing data in an automated system} \label{Modeling the process of obtaining and analyzing data in an automated system}

Let’s simulate obtaining and analyzing data in an automated system. Let’s imagine the
model of receiving and analyzing data in an automated system as a chain of processes
going from the stage of receiving a request from the user to sending a response to the
user. First, the primary source of data from the end user is a user request to the
system, transmitted in one channel of communication with a personal virtual assistant,
whether it be a voice or text request. If communication takes place in a voice channel,
it must present the data for further algorithmic processing in a form understandable for
algorithmic support, in some machine representation. Speaking of such a machine
representation, we can mean the translation of information in the natural language of
the user into a textual or indicative form. Receiving information in this form from the
user, it becomes possible to respond to the user’s request -- to decide in an automated
system that the user wanted to request and provide him with information or provide a
service as part of a business service. For example, if a user requested from a personal
virtual assistant an action from a business system in natural language, showing a
specific action (“turn on”, “say”, “I have a question about ...”), the personal virtual
assistant must answer the request or perform an action (turn on a specific service,
answer a question, delegate decision making to another business system).

The event chain of processes for processing a user’s request by a personal virtual assistant, which shows the stages of applying algorithmic support to analyze the request, is shown in Fig.~\ref{image-processing}.

\begin{figure*}[t!]
	\centerline{\includegraphics[width=\textwidth]{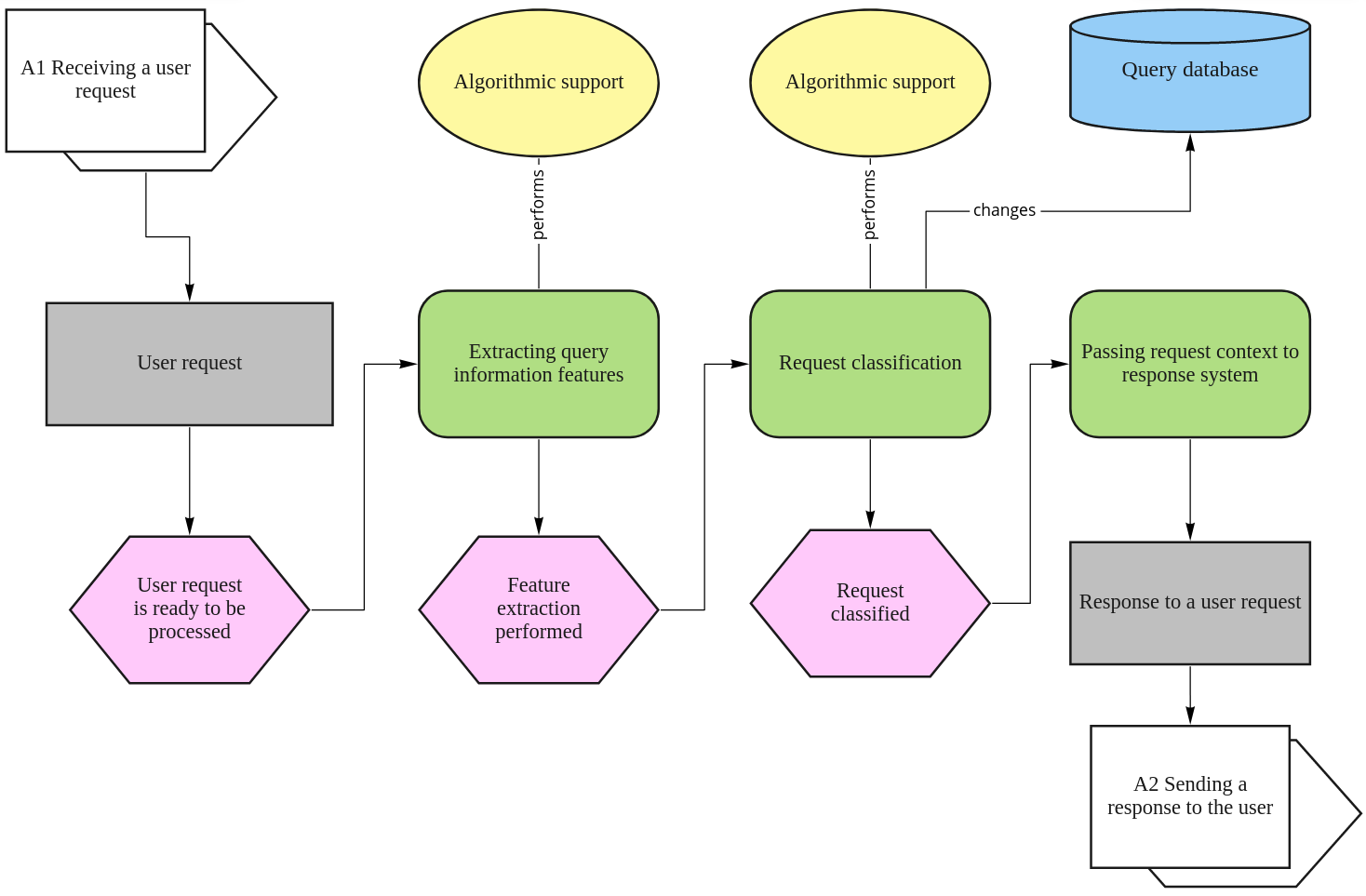}}
	\caption{The event chain of processes for processing a user's request by a personal virtual assistant.}
	\label{image-processing}
\end{figure*}

\section{Formation of a set of virtual assistant prototype technologies} \label{Formation of a set of virtual assistant prototype technologies}



\subsection{Selection of software components for algorithmic support} \label{Selection of software components for algorithmic support}

The connecting link of the components that underlie the personal virtual assistant will
be the programming language. Depending on the business service, the load and speed of
the system play an important role, so most times it is worth considering the presented
programming language options. In this article, the choice is made in favor of the Python
programming language, since over the past decade it has proven to be effective,
accelerating the creation of systems because of its syntax and a common solution for
creating algorithmic support in many scientific and applied fields, as well as for
creating software, in including for business \cite{c06}. An important factor in choosing
this language is many of open-source software libraries accepted by the professional
community for implementing algorithmic support based on approaches from the field of
machine learning \cite{c07}.

Speaking of open-source software libraries, there are some favorites among the libraries
accepted by the professional community for implementing machine learning algorithms.
Many large companies use the scikit-learn library to create software that uses machine
learning algorithms and is also used in the scientific community for research and
experimentation \cite{c08}. When working with user requests, it becomes possible to
build algorithmic support that uses information about the request, for example, by
processing the user’s natural speech in a textual representation. As an open-source
library for working with natural language, we will use the fastText library developed by
Facebook \cite{c09}. In the set of technologies, it is necessary to formalize the format
for receiving and transmitting data, as well as the possibility of transferring data to
the database. These can be generally accepted and common formats for storing and
transmitting JSON or XML data, however, the choice of formats is not limited and may
change based on speed or convenience reasons for storing data \cite{c10, c11}.

Such a set of technologies already makes it possible to create a prototype of a personal virtual assistant that can solve the tasks of processing client requests, being integrated into an automated request processing system. For further development of a prototype of a personal virtual assistant, it is necessary to formalize the interaction of algorithmic support components.

The scheme of interaction of the selected components is shown in Fig.~\ref{image-tools}.

\begin{figure*}[htbp]
	\centerline{\includegraphics[scale=0.3]{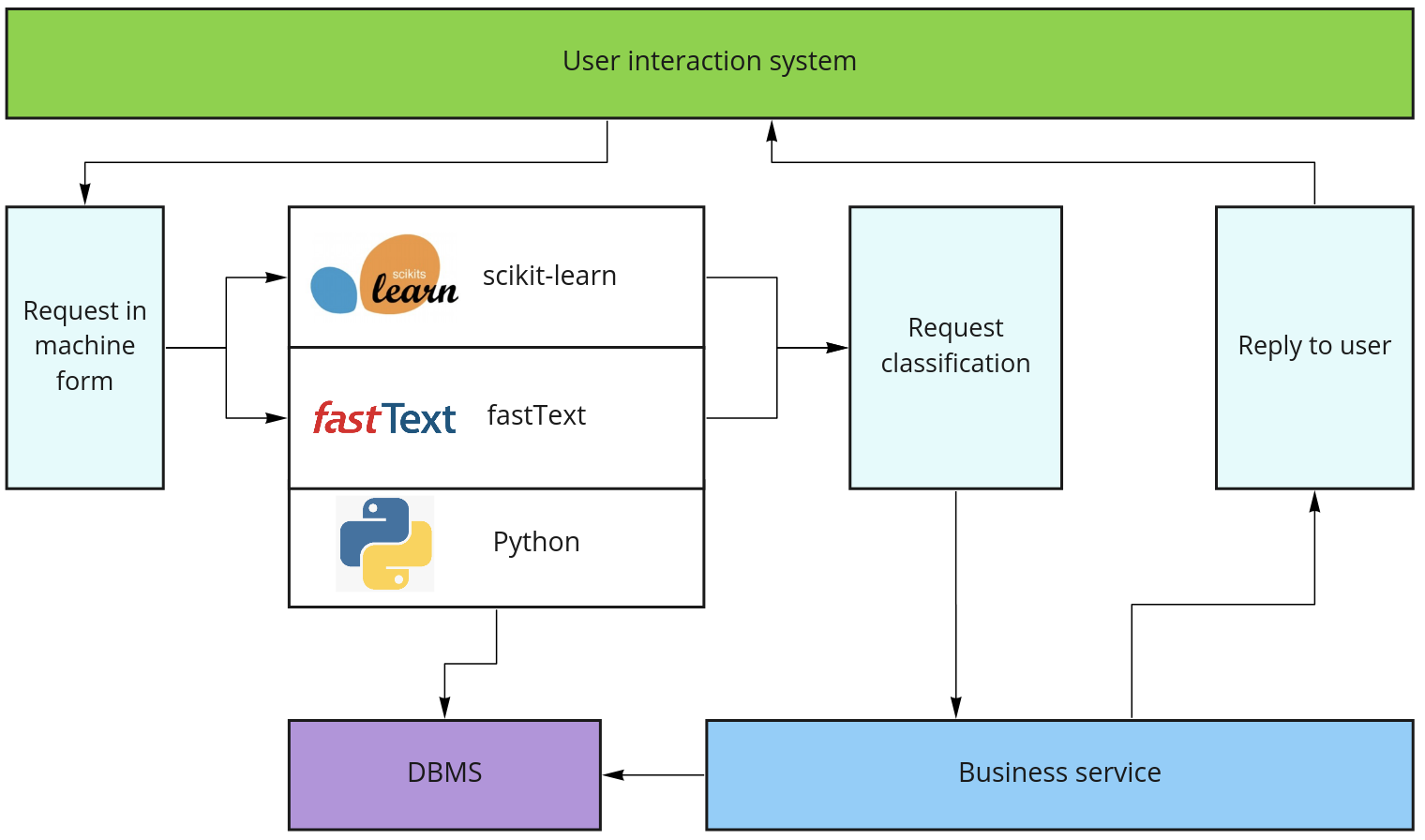}}
	\caption{The scheme of interaction of the selected components.}
	\label{image-tools}
\end{figure*}



\subsection{Development of software for algorithmic support} \label{Development of software for algorithmic support}

After selecting the algorithmic components, software was written in Python to load,
process, and classify natural language queries. The software is independent of the
hardware platform being run and can be used on both Windows and Linux.

\section{Results} \label{Results}

The results of the work described creating algorithmic support for the functioning of a
personal virtual assistant. We obtain the results of assessing the quality of the
classification of queries on The 20 Newsgroups data set \cite{c12}. When running the two
machine learning models, the weighted average value of the F1 metric was $82.5\%$. Fig.
\ref{table-results} show the results of classification in The 20 Newsgroups data set. It
also obtained the running time of machine learning models. The running time of the
logistic regression models of $189$ ms and the fastText model of $214$ ms. These results
show the applicability of the selected models in an industrial environment.

\begin{table}
	\caption{Results of classification in The 20 Newsgroups data set}
	\centering
	\begin{tabular}{|c|c|c|c|c|}
		\hline
		\textbf{class}        & \textbf{precision} & \textbf{recall} & \textbf{f1-score} & \textbf{support} \\
		\hline
		0                     & 0.752              & 0.734           & 0.743             & 319              \\
		1                     & 0.680              & 0.781           & 0.727             & 389              \\
		2                     & 0.740              & 0.731           & 0.736             & 394              \\
		3                     & 0.716              & 0.735           & 0.725             & 392              \\
		4                     & 0.828              & 0.816           & 0.822             & 385              \\
		5                     & 0.845              & 0.729           & 0.783             & 395              \\
		6                     & 0.764              & 0.897           & 0.825             & 390              \\
		7                     & 0.905              & 0.891           & 0.898             & 396              \\
		8                     & 0.940              & 0.945           & 0.942             & 398              \\
		9                     & 0.881              & 0.932           & 0.906             & 397              \\
		10                    & 0.947              & 0.945           & 0.946             & 399              \\
		11                    & 0.931              & 0.889           & 0.910             & 396              \\
		12                    & 0.764              & 0.784           & 0.774             & 393              \\
		13                    & 0.895              & 0.838           & 0.866             & 396              \\
		14                    & 0.898              & 0.919           & 0.908             & 394              \\
		15                    & 0.788              & 0.925           & 0.851             & 398              \\
		16                    & 0.711              & 0.901           & 0.795             & 364              \\
		17                    & 0.964              & 0.854           & 0.906             & 376              \\
		18                    & 0.780              & 0.594           & 0.674             & 310              \\
		19                    & 0.814              & 0.470           & 0.596             & 251              \\
		\hline
		\textbf{accuracy}     & 0.825              & 0.825           & 0.825             & 0                \\
		\textbf{macro avg}    & 0.827              & 0.815           & 0.817             & 7532             \\
		\textbf{weighted avg} & 0.829              & 0.825           & 0.823             & 7532             \\
		\hline
	\end{tabular}
	\label{table-results}
\end{table}

\vspace{12pt}

\end{document}